\documentclass[preprint,aps,superscriptaddress,preprintnumbers,nofootinbib]{revtex4}

\usepackage{inputenc}

\usepackage{graphicx}
\usepackage{graphics}
\usepackage{dcolumn}
\usepackage{bm}
\usepackage{color}
\usepackage{textcomp}
\usepackage{amsmath}
\usepackage{amssymb}
\usepackage{braket}

\usepackage{multirow}
\usepackage{setspace}
\usepackage[normalem]{ulem}

\usepackage{xcolor}

\usepackage{hyperref}
\hypersetup{
    colorlinks=true,       
    linkcolor=blue,        
    citecolor=blue,        
    filecolor=blue,        
    urlcolor=blue,         
    runcolor=blue
}

\newcommand{\meanval}[1]{\left\langle #1 \right\rangle}
\newcommand{\meanvalh}[1]{\left\langle \hat{ #1 } \right\rangle}

\newcommand{\om}[1]{\omega_{\mathrm{ #1 }}}
\newcommand{\ic}{\dot{\imath}}

\begin{document}

\title{Coherent anharmonicity transfer from matter to light in the THz regime}

\author{Mauricio Arias}
\affiliation{Departament of Physics, Facultad de Ciencias F\'isicas y Matem\'aticas, Universidad de Concepci\'on, Concepci\'{o}n, Chile}

\author{Johan F. Triana}
\affiliation{Departament of Physics, Universidad Cat\'olica del Norte, Av. Angamos 0610, Antofagasta, Chile}

\author{Aldo Delgado}
\affiliation{Departament of Physics, Facultad de Ciencias F\'isicas y Matem\'aticas, Universidad de Concepci\'on, Concepci\'{o}n, Chile}
\affiliation{ANID-Millennium Institute for Research in Optics, Chile}

\author{Felipe Herrera}\email{felipe.herrera.u@usach.cl}
\affiliation{ANID-Millennium Institute for Research in Optics, Chile}
\affiliation{Departament of Physics, Universidad de Santiago de Chile, Av. Victor Jara 3493, Santiago, Chile}

\date{\today}

\begin{abstract}
Optical nonlinearities are fundamental in several types of optical information processing protocols.
However, the high laser intensities needed for implementing phase nonlinearities using conventional optical materials represent a challenge for nonlinear optics in the few-photon regime.
We introduce an infrared cavity quantum electrodynamics (QED) approach for imprinting nonlinear phase shifts on individual THz pulses in reflection setups, conditional on the input power. 
Power-dependent phase shifts on the order of $ 0.1\, \pi$ can be achieved with femtosecond pulses of only a few $\mu$W input power. 
The proposed scheme involves a small number of intersubband quantum well transition dipoles evanescently coupled to the near field of an infrared resonator. 
The field evolution is nonlinear due to the dynamical transfer of spectral anharmonicity from material dipoles to the infrared vacuum,  through an effective dipolar chirping mechanism that transiently detunes the quantum well transitions from the vacuum field, leading to photon blockade. 
We develop analytical theory that describes the dependence of the imprinted nonlinear phase shift on relevant physical parameters. 
For a pair of quantum well dipoles, the phase control scheme is shown to be robust with respect to inhomogeneities in the dipole transition frequencies and relaxation rates. 
Numerical results based on the Lindblad quantum master equation validate the theory in the regime where the material dipoles are populated up to the second excitation manifold. 
In contrast with conventional QED schemes for phase control that require  strong light-matter interaction, the proposed phase nonlinearity works best in weak coupling, increasing the prospects for its experimental realization using current nanophotonic technology.  
\end{abstract}

\maketitle

\vspace{3cm}
\section{Introduction}

Cavity quantum electrodynamics is one of the building blocks of quantum technology \cite{Haroche2020,OBrien2009}. 
Strong light-matter interaction between dipolar material resonances and the electromagnetic vacuum of a cavity has been used for protecting and manipulating quantum information across the entire frequency spectrum using neutral atoms \cite{Kimble:1998,Ruddell2017}, semiconductors \cite{Mabuchi2002,Vahala:2003wj,Ballarini:2019um,Kiraz:2003ty}, and superconducting artificial atoms \cite{Blais2004,Fink:2008wp,Putz2014,Reagor2016,Blais:2020vz}.
Cavity QED observables such as the vacuum Rabi splitting have also been demonstrated with material dipoles in infrared (THz) resonators at room temperature using intersubband transitions \cite{Dini2003,Gunter:2009,Mann2021,Paul:2023tt} and molecular vibrations \cite{Long2015,Dunkelberger2016,Xiang2019,George2016,George2015,Shalabney2015coherent}, for applications such as infrared photodetection \cite{Vigneron:2019} and controlled chemistry \cite{Nagarajan:2021ur,Ahn2023}.  
The enhancement of the spontaneous emission rate of material dipoles in a weakly coupled cavity via the Purcell effect  \cite{Milonni:1973wf,Barnes1998,Akselrod:2014} has been used over different frequency regimes for reservoir engineering \cite{Harrington2022,Jacob:2012vh}, dipole cooling \cite{Genes2008,Zambon2022} and quantum state preparation \cite{Petiziol2022}. 
In infrared cavities, the Purcell effect can be an effective tool for studying the relaxation dynamics of THz transitions in materials \cite{Nishida2022,herrera2022,Metzger2019}, given the negligible radiative decay rates at these frequencies in comparison with non-radiative relaxation processes \cite{Vilas2023,Lyubomirsky:1998}. 
The direct linear measurement of confined infrared field dynamics in a weakly coupled dipole-cavity system \cite{Wilcken2023} can thus provide information about infrared transitions that otherwise would only be accessible using ultrasfast nonlinear spectroscopy \cite{Golonzka2001,Anfuso:2012wy}.

Cavity QED also enables the manipulation of external electromagnetic fields that drive a coupled cavity-dipole system \cite{Mabuchi2002}. 
Implementing conditional phase shifts via intracavity light-matter interaction can be used for quantum information processing \cite{Zubairy2003,Duan2004,Enk:2004ws}, as demonstrated using atomic dipoles \cite{Turchette1995,Tiecke:2014} and quantum dots \cite{Young2011,Hughes2012} in optical cavities. 
In analogy with classical phase modulation processes in bulk nonlinear optical materials \cite{Ho1991}, which depend on the anharmonic response of the medium to strong  fields \cite{Axt1998}, cavity-assisted phase shifts are possible due to photon blockade effects that arise due to intrinsic spectral anharmonicities of strongly coupled light-matter systems \cite{Faraon2010,Birnbaum2005}. 
In the infrared regime, despite the growing interest in cavity QED phenomena with molecular vibrations \cite{Nagarajan:2021ur,Xiang2018,Herrera2020perspective,Grafton2020,Kadyan2021,Ahn2023,Wilcken2023,Wright:2023,Menghrajani:2022tr} and semiconductors \cite{Zaks_2011,Autore2018,Mann2021,Bylinkin:2021}, viable physical mechanisms for implementing conditional phase dynamics with infrared fields have yet to be developed. 

Here we study a previously unexplored form of dynamical photon blockade effect that can be used for imprinting intensity-dependent phase shifts on electromagnetic pulses at THz frequencies (mid-infrared). 
The physical mechanism that supports the phase nonlinearity involves an effective transfer of the spectral anharmonicity of suitable few-level systems to the near field of an infrared resonator in weak coupling. 
To emphasize the feasibility of implementing the proposed phase nonlinearity using current technology, the relevant frequency scales of the problem are specified according to recent cavity QED experiments with intersubband transitions of multi quantum wells (MQWs) embedded in infrared nanoresonators \cite{Mann2021}. 
By driving the resonator with a moderately strong femtosecond laser pulse, the matter-induced phase nonlinearity can be retrieved from the free-induction decay (FID) of the resonator near field $E_{\rm nf}(t)$ using linear heterodyne spectroscopy techniques with femtosecond time resolution \cite{Muller2018,Xu2013,Metzger2019,Nishida2022,Pollard2014,Autore2021,Huth2013}. 
The article is organized as follows: Section \ref{sec:dynamics} describes the model for a MQW coupled to a common open cavity field.
Section \ref{sec:chirping} develops a mean-field theory of the dynamical chirping effect that gives rise to the infrared phase nonlinearity. 
Sec. \ref{sec:phaseshift} discusses the scaling of the predicted nonlinear phase shift with the physical parameters of the problem for identical intersubband dipoles. 
Sec.  \ref{sec:darksystem} shows that the results are robust with respect to dipole inhomogeneities. 
Numerical validation of the theory is given in Sec. \ref{sec:validation} using a Lindblad quantum master equation description of the system dynamics. 
We summarize and discuss perspectives of this work in the Conclusions.
%

\section{Cavity QED model of anharmonic semiconductor dipoles}
\label{sec:dynamics}

\begin{figure*}[t!]
\includegraphics[width=0.55\textwidth]{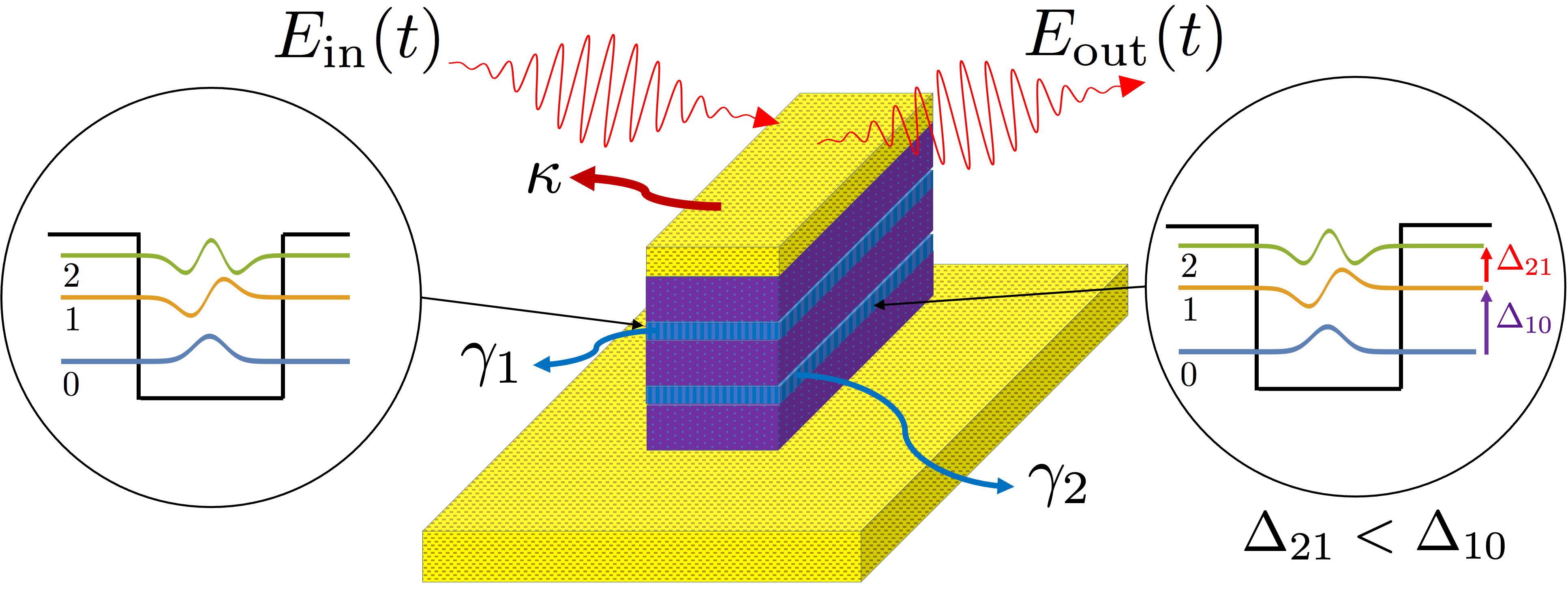}
\caption{{\bf Schematic picture of a MQW in a driven open infrared nanocavity}. 
Two quantum wells forming a MQW with fundamental frequency $\omega_0=\Delta_{10}$. 
The $0\to1$ and $1\to2$ transitions are coupled to the near-field of an open nanocavity with frequency $\omega_{\mathrm{c}}$ by an incoming laser pulse $E_{\mathrm{in}}$.
%
%
The photons that leak out the nanocavity on sub-picosecond time scales at rate $\kappa$ generate the free-induction decay (FID) signal encoded in the outcoming field $E_{\mathrm{out}}$. 
} 
\label{fig:scheme}
\end{figure*}

%
We consider a small number $N$ of quantum wells located within the near field of a resonant infrared nanoantenna, as illustrated in Fig. \ref{fig:scheme}. The quantum wells do not interact with each other and have discrete intersubband energy levels with transition frequencies in the THz regime \cite{Oppo1997}.
The bare Hamiltonian of the MQW system is a collection of anharmonic quantum Kerr oscillators  ( $\hbar \equiv 1$ throughout)
\begin{equation}\label{eq:locH}
\hat{H}_n = \omega_n \hat{b}_n^\dagger \hat{b}_n - U_n \hat{b}_n^\dagger \hat{b}_n^\dagger \hat{b}_n\hat{b}_n,
\end{equation}
where $\hat{b}_n$ is the annihilation operator of the $n$-th quantum well dipole, $\omega_n$ is the fundamental frequency and $U_n$ the anharmonicity parameter. 
The geometry of the quantum well structure determines the confined charge carriers potential and the spectral anharmonicity \cite{Goulain2023}.
%
%
 
%
Projecting into a complete eigenbasis $|\nu_{n}\rangle$, Eq. (\ref{eq:locH}) can be written as 
%
$\hat{H}_n = \sum_\nu E_{\nu_{n}} |\nu_{n}\rangle\langle \nu_{n}|$, 
%
with eigenvalues $E_{\nu_{n}} = \omega_n\nu_{n} - U_n(\nu_{n}^{2}-\nu_{n}) $. 
The energy difference between consecutive levels of the $n$-th quantum well is $E_{\nu+1} - E_\nu = \omega_n - 2U_n\nu$. 
The nonlinear parameter $U_n$ thus lowers the energy spacing of the $1 \rightarrow 2$ excitation by $\Delta=2U_n$ relative to the fundamental frequency. 
In comparison with molecular vibrations, for which $\Delta \sim 10 - 40$ cm$^{-1}$ \citep{Fulmer2004,Dunkelberger2019}, multi-quantum well dipoles enable much larger anharmonicities, with $U_n\sim 100-300$ cm$^{-1}$ \citep{Mann2021}. 

The anharmonic quantum wells couple to a common resonator near field $\hat{a}$, as described by the Hamiltonian 
\begin{equation}\label{eq:H1}
\hat{\mathcal{H}} = \omega_{\rm c} \hat{a}^\dagger\hat{a} + \sum_{n=1}^N \left[\hat{H}_{n} + g_n(\hat{a}\hat{b}^{\dagger}_{n} + \hat{a}^{\dagger}\hat{b}_{n})\right],
\end{equation}
where $\omega_{\rm c}$ is the resonant field mode frequency and $g_{n}=\mathcal{E}_{0}d_{0}$ is the light-matter coupling strength. 
The latter depends on the square-root amplitude of the vacuum fluctuations $\mathcal{E}_{0}$ and the transition dipole moment $d_{0}$. 
For simplicity, we assume transition dipoles are state-independent but other choices do not qualitatively affect the results.

We calculate the dissipative dynamics of the system in the presence of a driving pulse according to the Lindblad quantum master equation
\begin{equation}\label{eq:lindblad}
\frac{\rm d}{\rm dt}\hat{\rho} = -\ic[\hat{\mathcal{H}} + \hat{H}_{\rm d}(t),\hat{\rho}] + \mathcal{L}_\kappa[\hat{\rho}]+\mathcal{L}_{\gamma}[\hat{\rho}].
\end{equation}
where $\hat{\rho}$ is the density matrix of the total cavity-MQW system. 
$\mathcal{L}_{\kappa}[\hat{\rho}]$ and $\mathcal{L}_{\gamma}[\hat{\rho}]$ are photonic and material relaxation superoperators given by 
\begin{eqnarray}
\mathcal{L}_\kappa[\hat \rho] &=& \frac{\kappa}{2}\left(2\,\hat a\hat \rho\hat a^\dagger - \hat a^\dagger \hat a\, \hat \rho - \hat \rho\,\hat a^\dagger\hat a \right),\label{eq:L kappa}\\
\mathcal{L}_{\gamma}[\hat \rho] &=&\sum_{n=1}^N \frac{\gamma_n}{2}\left(2\,\hat b_n\hat \rho\hat b_n^\dagger - \hat b_n^\dagger \hat b_n\, \hat \rho - \hat \rho\,\hat b_n^\dagger\hat b_n \right).\label{eq:L local}
\end{eqnarray}
where $\kappa$ and $\gamma_n$ are the decay rates of photonic and $n$-th quantum well modes, respectively.
The decoherence processes encoded in $\gamma_n$ are mainly given by the interaction between MQW and the thermalized phonon bath of the semiconductor structure \citep{Levine1993}. For the open cavity field, the main source of decoherence is non radiative decay in the metal \citep{Mann2020}, as well as radiative losses \citep{herrera2022}.
The time-dependent Hamiltonian $\hat{H}_{\rm d}(t)$ that describes the driving pulse is given by 
\begin{equation}\label{eq:driving}
\hat{H}_{\rm d}(t) = F_0\varphi(t)(\hat{a}e^{i\omega_{\rm d} t} + \hat{a}^\dagger e^{-i\omega_{\rm d} t}),
\end{equation}
with the Gaussian pulse envelope $\varphi(t) = \exp[-(t-t_0)^2/(2T^2)]$ and carrier frequency $\omega_{\rm d}$. $|F_0|^2$ is proportional to the incoming photon flux \footnote{The steady photon flux in the continous wave regime of an empty cavity with radiative decay rate $\kappa$ is $\Phi_{\rm flux}=\kappa\langle\hat{a}^\dagger\hat{a}\rangle=4|F_0|^2/\kappa$.}, $t_0$ is the pulse center time and and $T$ is the pulse duration.


\section{Mean-field Nonlinear chirping model}
\label{sec:chirping}

\begin{figure*}[t!]
\includegraphics[width=0.995\textwidth]{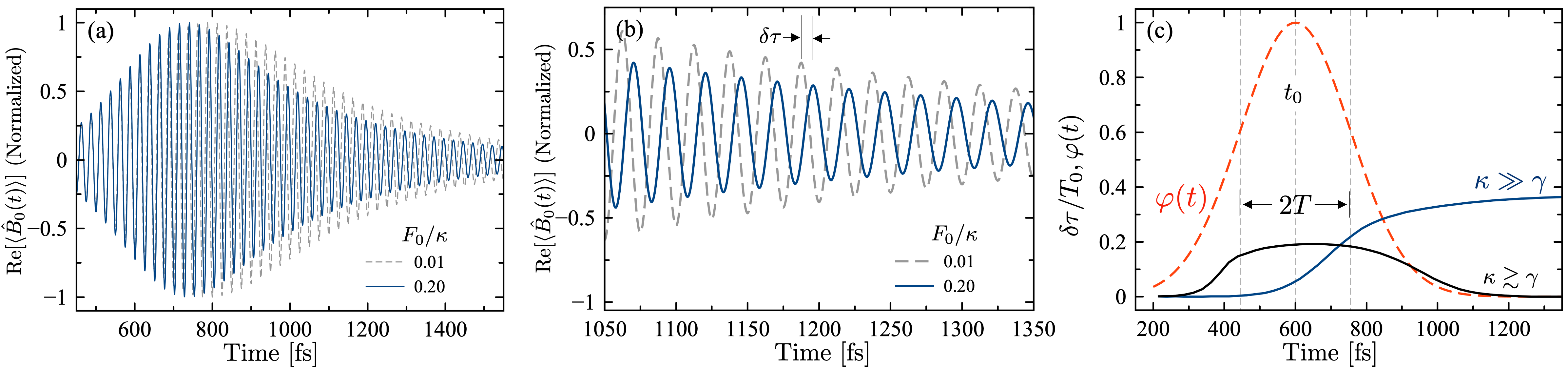}
\caption{{\bf Dipolar FID signals.} (a) Real part of the material collective coherence $\text{Re}\langle\hat{B}_0(t)\rangle$ for two identical quantum wells coupled to a common field mode. The antenna is driven by a laser pulse with a carrier frequency $\omega_{\rm d}$, pulse duration $T=155$ fs and centered at $t_{0}=600$ fs for laser intensities $F_0=0.01\kappa$ and $F_0=0.20\kappa$. Additional parameters are $\{\omega_{0},\kappa,\gamma,U,\sqrt{N}g\}=\{40.0,12.0,0.6,0.6,1.0\}$ THz. 
(b) Zoom of a fraction of the FID signals shown in (a) to visualize the time delay $\delta\tau$ generated in $t>t_{0}+2T$. (c) Time delay ratio calculated as the time difference between peaks and dips of the material collective coherence for $\gamma=0.6\text{ THz}$ (blue curve, different material and nanocavity bandwidths) and $\gamma=10.0\text{ THz}$ (black curve, comparable bandwidths). Dashed red line corresponds to the Gaussian pulse envelope. 
} 
\label{fig:timedomain}
\end{figure*}

We assume that the system dynamics involves only the lowest three energy levels of the quantum wells (i.e. $\nu_{\rm max}=2$). To ensure that higher energy levels do not participate in the evolution, we assume weak driving conditions, $F_0/\kappa<1$. 
Mean-field equations of motion for identical quantum wells can be obtained from Eq. (\ref{eq:lindblad}) for light and matter coherences, to give coupled non-linear system
\begin{align}
\label{eq:acoh}
\frac{\rm d}{\rm dt}\meanvalh{a} & = -\left(\frac{\kappa}{2}+i\omega_{\rm c}\right)\meanvalh{a} - i\sqrt{N}g\langle\hat{B}_0\rangle - i\tilde{F}_{\rm d}(t)\\ 
\label{eq:bcoh}
\frac{\rm d}{\rm dt}\langle\hat{B}_0\rangle  & =- \left[\frac{\gamma}{2}+i\left(\omega_0-\frac{2U}{N}|\langle\hat{B}_0\rangle|^2\right)\right]\langle\hat{B}_0\rangle - i\sqrt{N}g\langle\hat{a}\rangle,
\end{align}
where $\hat{B}_0 = (1/\sqrt{N})\sum_{n}\hat{b}_{n}$ is the bright collective oscillator mode of fundamental frequency $\omega_{0}$ and decay rate $\gamma$. The driving parameter is $\tilde{F}_{\mathrm{d}}=F_{0}\varphi(t)\exp(-i\omega_{\mathrm{d}}t)$. For simplicity we set $g_{n}=g$ and $U_{n}=U$.
For homogeneous systems, the $N-1$ dark collective modes $\hat{B}_{\alpha} = (1/\sqrt{N})\sum_{n}c_{\alpha,n}\hat{b}_{n}$, with $c_{\alpha,n}=\exp(i2\alpha n/N)$, evolve completely decoupled from Eqs. (\ref{eq:acoh}) and (\ref{eq:bcoh}) (see Appendix \ref{app:anharmonic}).
%
The Kerr nonlinearity generates an effective dipole chirping effect, with instantaneous frequency 
\begin{equation}
\omega'_0(t) = \omega_0 - \frac{2U}{N}|\langle \hat{B}_0(t)\rangle|^2. 
\label{eq:effectivefrequency}
\end{equation} 
This is red shifted from the fundamental resonance by an amount proportional to the bright mode occupation. 
The nonlinearity is proportional to the anharmonicity parameter $U$ and is small for large $N$ \cite{herrera2022}. The transient red shift occurs while the system is driven by the laser pulse, which populates $\hat{B}_0(t)$, and is thus proportional to the photon flux parameter $F_{0}$.

We solve Eqs. (\ref{eq:acoh}) and (\ref{eq:bcoh}) analytically to gain insight on the chirping effect.
We assume that the bandwidth of the dipole resonance is much smaller than the antenna bandwidth, i.e., $\kappa\gg\gamma$. By adiabaticaly eliminating the antenna field from the dynamics, the evolution of bright mode after the pulse is over is given by
\begin{align}
\label{eq:ampb0_fd0m}
\langle \hat{B}_{0}(t)\rangle & = B_{\rm off} e^{-\frac{\tilde{\gamma}}{2}(t-t_{\rm off})}e^{i\phi(t)} 
\end{align}
where $t_{\mathrm{off}}$ is the pulse turn-off time. 
The phase evolves as
\begin{align}
\label{eq:ampphi_fd0m}
\phi(t) & = \phi_{\rm off} +\frac{2UB_{\rm off}^2}{N\tilde{\gamma}}\left\{1 - e^{-\tilde{\gamma}(t-t_{\rm off})}
\right\},
\end{align}
where $\tilde{\gamma} =\gamma(1+4Ng^{2}/\kappa\gamma)$ is the Purcell-enhanced dipole decay rate \cite{Plankensteiner2019,Metzger2019} and $B_{\rm off}=|\langle \hat{B}_{0}(t_{\rm off})\rangle|$.
Defining $\tau=t-t_{\rm off}$, in the long time regime, $\tau\tilde{\gamma} \gg 1$, Eq. (\ref{eq:ampphi_fd0m}) gives the stationary relative phase
\begin{equation}
\Delta\phi_{\rm ss} = \phi_{\rm ss} - \phi_{\rm off} = \frac{2UB_{\rm off}^2}{N\tilde{\gamma}}, 
\label{eq:ssphase}
\end{equation}
which depends quadratically on the laser strength, through the implicit linear dependence of $B_{\rm off}$ on $F_{0}$.
The derivation of Eq. (\ref{eq:ampphi_fd0m}) can be found in Appendix \ref{app:adibaticelimination}.
In the limiting cases of harmonic oscillators ($U=0$), thermodynamic limit ($N\to\infty$), or linear response ($F_{0}/\kappa\ll 1$), the relative phase is neglegible $(\Delta\phi_{\mathrm{ss}}\approx0)$. 
Molecular ensembles have low anharmonicites, and have been shown to require higher pulse strengths to produce finite relative phases \cite{herrera2022} than the ones discussed here.

Figures \ref{fig:timedomain}a and \ref{fig:timedomain}b show the evolution of the dipole coherence $\mathrm{Re}[\langle\hat{B}_{0}(t)\rangle]$ obtained by solving Eqs. (\ref{eq:acoh})-(\ref{eq:bcoh}) numerically with parameters relevant for experimental implementations \cite{Mann2021,Goulain2023}. 
Figure \ref{fig:timedomain}b shows the time delay $\delta\tau$ produced by a strong driving pulse ($F_{0}/\kappa=0.2$) on the FID signal, in comparison with weak pulses.
The delay $\delta\tau$ in time domain results in the relative phase from Eq. (\ref{eq:ssphase}).  
Recent experiments with infrared nanoantennas have the femtosecond temporal resolution necessary to measure $\delta\tau$ \cite{Metzger2019,herrera2022}.

Figure \ref{fig:timedomain}c shows the evolution of $\delta\tau$ for two scenarios. 
For long dipole lifetimes ($\gamma\ll\kappa$), i.e., narrowband MQW response, the time delay of the FID signal remains after the driving pulse is over. On the contrary, when $\gamma\sim\kappa$ the time delay disappears after the pulse ends. 
The system thus requires long dipole dephasing times to imprint a stationary time delay in the near field once the driving pulse is turned off.
%


\section{Nonlinear phase shift in the frequency domain}
\label{sec:phaseshift}

\begin{figure*}[t!]
\includegraphics[width=0.95\textwidth]{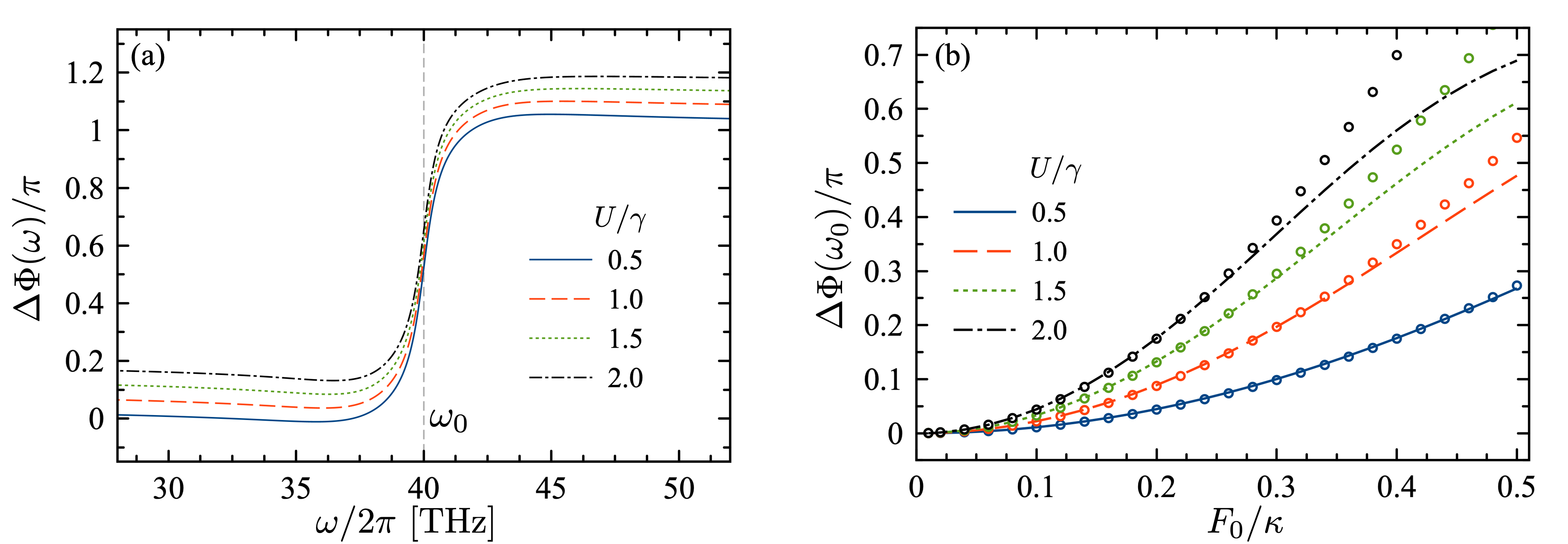}
\caption{{\bf Intensity-dependent phase spectrum of the scattered field.} (a) Relative phase spectrum $\Delta\Phi(\omega)$ for $F_{0}=0.2\kappa$ and different values of $U/\gamma$. (b) Nonlinear phase spectrum $\Delta\Phi(\omega_0)/\pi$ at $\omega_0$ as a function of laser strength parameter $F_{0}/\kappa$. Circles correspond to analytical quadratic fitting with fit parameter $\alpha=3.5$ for all $U/\gamma$ cases . Both panels are calculated using Eq. (\ref{eq:phasespectrum}) for different ratios $U/\gamma$. Additional parameters are $\{\omega_{0},\kappa,\gamma,\sqrt{N}g\}=\{40.0,12.0,0.6,1.0\}$ THz.
} 
\label{fig:freqdomain}
\end{figure*}

The time delay in the collective material coherence $\langle\hat{B}_{0}(t)\rangle$ from Fig. \ref{fig:timedomain} is transferred to the photonic coherence $\langle\hat{a}(t)\rangle$, which ultimately gives the observable FID signal in heterodyne measurements \cite{herrera2022,Metzger2019}. 
We define the Fourier transform of the field coherence as
\begin{equation}
\langle\hat{a}(\omega)\rangle = \frac{1}{\sqrt{2\pi}}\int_{-\infty}^{\infty} {\rm d}t \langle\hat{a}(t)\rangle e^{i\omega t},
\end{equation}
and calculate the phase response $\Phi(\omega)$ of the FID signal as
\begin{equation}
\label{eq:phasespectrum}
\Phi(\omega) = \arctan\left(\frac{\text{Im}[\langle\hat{a}(\omega)\rangle]}{\text{Re}[\langle \hat{a}(\omega) \rangle]}\right).
\end{equation}
The Fourier transform is taken for the post-pulse FID signal.
Fig. \ref{fig:freqdomain}a shows the phase spectrum for different values of the parameter $U/\gamma$, at fixed driving strength $(F_0=0.2\kappa)$.
The relative phase in Fourier space is negligible for the limiting cases discussed above. 
In the case of anharmonic MQWs, the relative phase $\Delta\Phi(\omega)=\Phi(\omega)-\Phi_{\rm harm}$ 
increases as the anharmonicity parameter $U/\gamma$ grows for fixed $F_{0}$, with  $\Phi_{\rm harm}$ given by relative phase obtained under harmonic conditions.

To gain insight on the behavior of the relative nonlinear phase, we use Eq. (\ref{eq:ampb0_fd0m}) to calculate $\Delta\Phi(\omega)$ analytically in the Fourier domain. We obtain
\begin{equation}
\label{eq:analyticalphase}
\Delta\Phi(\omega_0) =\alpha  \frac{2U}{N\tilde{\gamma}}\left( \frac{F_0}{\kappa} \right)^{2},
\end{equation}
where the numerical parameter $\alpha$ comes from the proportionality relation $|\langle\hat{B}_{0}(t)\rangle|\propto F_{0}$ and the fact that $\phi_{\rm off}$ in Eq. (\ref{eq:ampphi_fd0m}) is also a nonlinear phase that grows with the dipole amplitude before the driving pulse is off. The derivation of Eq. (\ref{eq:analyticalphase}) and additional details can be found in Appendix \ref{app:adibaticelimination}.
Figure \ref{fig:freqdomain}b shows the analytical fitting and numerical calculations for the nonlinear phase shift at the fundamental frequency $\omega_{0}$ as a function of the laser intensity parameter $F_{0}/\kappa$, for different values of $U/\gamma$. 
In the low anharmonicity regime $U/\gamma<1$, the nonlinear phase has a clear quadratic dependence on $F_{0}$, while for strong anharmonicities $U/\gamma\gtrsim1$, $\Delta\Phi(\omega_{0})$ is quadratic for only up to a certain driving strength.
Beyond this point, higher energy levels ($\nu>2$) start to contribute with the dynamics of the system and the adiabatic elimination approach used to derive Eq. (\ref{eq:analyticalphase}) breaks down.


\section{Nonlinear phase shift enhancement via dark states}
\label{sec:darksystem}

The dynamics in the totally symmetric case with identical QW dipoles only involves the field and bright collective modes without the influence of the dark manifold. 
For a pair of inhomogeneous QWs ($N=2$), the equation of motion for the bright mode $\langle\hat{B}_{0}\rangle$ should be extended to read
\begin{align}
\label{eq:brightmode}\frac{\rm d}{\rm dt}\langle \hat{B}_0 \rangle & =  -\left(\frac{\bar{\gamma}}{2}+i\bar{\omega}(t)\right)\langle \hat{B}_0 \rangle - \left(
\frac{\Delta\gamma}{2} + i\Delta\omega(t)
\right)\langle \hat{B}_1 \rangle\nonumber\\
& \quad - i \sqrt{N}g\langle \hat{a}\rangle
\end{align}
where $\langle \hat{B}_1 \rangle=(-\langle \hat{b}_1\rangle + \langle \hat{b}_2\rangle)/\sqrt{2}$ is the dark mode for the pair of dipoles, 
$\bar{\gamma}=(\gamma_1+\gamma_2)/2$ and $\Delta\gamma = (\gamma_2 - \gamma_1)/2$ are the average value and the mismatch of decay rates, respectively.
The instantaneous average frequency is now given by 
\begin{align}
\bar{\omega}(t) & = \bar{\omega} - U\left(|\langle \hat{B}_0(t)\rangle|^2+|\langle \hat{B}_1 (t)\rangle|^2\right)
\end{align}
and the frequency mismatch by 
\begin{align}
\Delta\omega(t) & = \Delta\omega  - 2U{\rm Re}\left[\langle \hat{B}_0(t)\rangle^*\langle \hat{B}_1(t)\rangle\right],
\end{align}
where $\bar{\omega} = (\omega_1+\omega_2)/2$ and $\Delta\omega = (\omega_2 - \omega_1)/2$. 
The coupling of the dark mode $\langle\hat{B}_{1}\rangle$ with the bright mode influences the dynamics of the system. 
It is clear that in the totally symmetric case ($\Delta\gamma=\Delta\omega=0$), bright and dark modes are decoupled and Eq. (\ref{eq:brightmode}) reduces to Eq. (\ref{eq:bcoh}).
The positive additive contribution $|\langle\hat{B}_{1}\rangle|^{2}$ to the instantaneous dipole frequency  suggests that  the dark manifold enhances the nonlinear phase shift.
The derivation of Eq. (\ref{eq:brightmode}) can be found in Appendix \ref{app:asymmetric}.

\begin{figure*}[t!]
\includegraphics[width=0.995\textwidth]{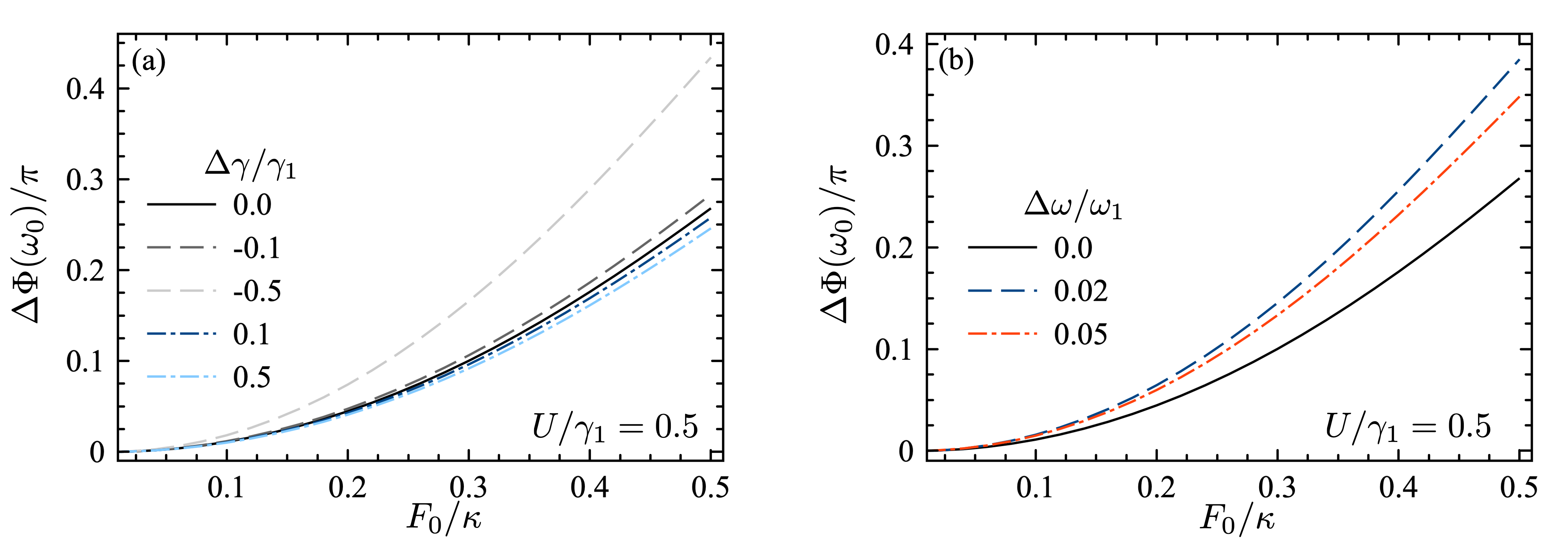}
\caption{{\bf Nonlinear phase shift enhancement via dipole inhomogeneity}. Nonlinear phase shift at $\omega_{0}$ as a function of the driving ratio $F_0/\kappa$ for the anharmonicity parameter $U/\gamma=0.5$. (a) For $\omega_{1}=\omega_{2}=\omega_{0}$ and $\gamma_{1}\neq\gamma_{2}$. (b) For $\omega_{0}=\omega_{1}\neq\omega_{2}$ and $\gamma_{1}=\gamma_{2}$. Additional parameters are $\{\omega_{0},\kappa,\gamma_{1},\sqrt{N}g\}=\{40.0,12.0,0.6,1.0\}$ THz.
} 
\label{fig:fluctuations}
\end{figure*}

Figure \ref{fig:fluctuations} shows the nonlinear phase spectrum of $\langle\hat{a}(\omega)\rangle$ evaluated at the fundamental frequency $\omega_{0}$ for two inhomogeneus scenarios. 
First, we set equal fundamental frequencies with variations in the decay rates of the QWs, i.e., $\omega_{1}=\omega_{2}=\omega_{0}$ and $\gamma_{1}\neq\gamma_{2}$. 
In Fig. \ref{fig:fluctuations}a we find both enhancement and reduction of the nonlinear phase for $U=0.5\gamma_{1}$. Enhancement is reached for $\Delta\gamma<0$, which is due to a reduction of the effective decay rate $\bar{\gamma}$ associated with bright mode in Eq. (\ref{eq:brightmode}). 
In the opposite case, i.e., for $\gamma_{2}>\gamma_{1}$, $\Delta\Phi(\omega_{0})$ decreases due to the increased decay rate of the bright mode. 
The second scenario considers one QW detuned with respect to the other but both having equal decay rates, i.e., $\omega_{1}\neq\omega_{2}$ and $\gamma_{1}=\gamma_{2}$. The results are shown in Fig. \ref{fig:fluctuations}b.
In this case, the small blue and red detuning of one of the quantum wells gives a nonlinear phase enhancement. 
However, for very large detunings ($\Delta\omega/\omega_{0}>0.1$) the nonlinear phase tends to the homogeneous limit ($\Delta\omega=0$), as the detuned quantum well becomes effectively decoupled from the field.


\section{Validity of the mean-field theory}
\label{sec:validation}

\begin{figure*}[t!]
\includegraphics[width=1\textwidth]{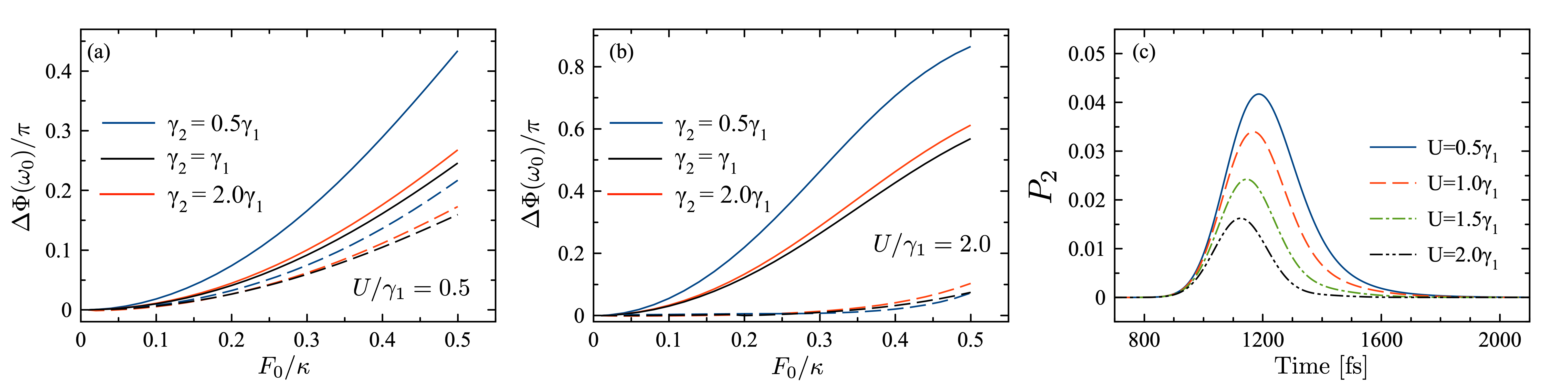}
\caption{{\bf Fully quantum validation of the nonlinear phase shift.} Relative nonlinear phase shift as a function of laser strength parameter $F_{0}/\kappa$ found in the mean-field approximation (solid lines) and by solving the Lindblad quantum master equation (dashed lines) for different inhomogeneous scenarios and for (a) $U/\gamma_{1}=0.5$ and (b) $U/\gamma_{1}=2.0$.  (c) Population dynamics of the second excited energy level $P_{2}$ for $F_{0}=0.3\kappa$ and different values of the anharmonicity parameter $U/\gamma_{1}$.  Additional parameters for all panels are $\{\omega_{0},\kappa,\gamma_{1},\sqrt{N}g\}=\{40.0,12.0,0.6,1.0\}$ THz.
} 
\label{fig:qmvalidation}
\end{figure*}

We calculate the evolution of the total density matrix by solving the Lindblad quantum master equation in Eq. (\ref{eq:lindblad}) to explore the regime of validity of the analytical theory for a pair of quantum well dipoles. 

Figures \ref{fig:qmvalidation}a and \ref{fig:qmvalidation}b show the nonlinear phase shift $\Delta\Phi(\omega_{0})$ as a function of the laser strength parameter $F_{0}/\kappa$ predicted by the Lindblad QME for two different values of the anharmonicity: $U=0.5\gamma_{1}$ and $U=2.0\gamma_{1}$. 
For $U=0.5\gamma_{1}$, both mean-field (solid lines) and fully quantum (dashed lines) calculations agree up to a factor of two approximately. 
However, $\Delta\Phi(\omega_{0})$ in the mean-field approach is overestimated for $U=2.0\gamma_{1}$, where the nonlinear phase is negligible up to $F_{0}/\kappa\approx 0.4$, then increases for higher laser intensities.
The latter is due to the large detuning between cavity frequency $\omega_{\mathrm{c}}$ and $\Delta_{21}$.
As a consequence, anharmonic blockade is observed for the intersubband state $\nu=2$ such that the higher level does not participate in the dynamics, causing the dipole to behave effectively as a harmonic oscillator. 
In Fig. \ref{fig:qmvalidation}c, we show the population of the $\nu=2$ level ($P_{2}$) for $F_{0}/\kappa=0.3$ and different values of the anharmonicity parameter $U/\gamma_{1}$.
$P_{2}$ is suppressed as $U$ increases due to the blockade effect. 
The mean-field theory is accurate for weak pulses ($F_{0}/\kappa<0.1$) independent on the anharmonicity parameter $U/\gamma$ and, for stronger pulses with anharmonicities lower than dissipation rate of MQWs ($U/\gamma<1.0$).

\section{Discussion and Conclusion}
\label{sec:conclusions}

In this work, we described a novel dynamical photon blockade mechanism in THz cavity QED that can be used for imprinting power-dependent phase shifts on the electromagnetic response of a coupled cavity-dipole system. 
We develop analytical quantum mechanical theory to model free-induction decay signals of a pulse-driven cavity system, using parameters that are relevant for quantum well intersubband transitions  in mid-infrared resonators \cite{Mann2021}. 
For $N$ quantum wells within the near field of the driven resonator, the theory shows that using only a moderately strong pulse that drives a small fraction of the intersubband level population to the second excitation manifold, a stationary phase shift proportional to the spectral anharmonicity parameter $U/N\gamma$ and the photon flux of the pulse, can be imprinted on the FID response of the near field, which can then be retrieved using time-domain spectroscopic techniques \cite{Wilcken2023}. 
For experimentally relevant system parameters, nonlinear phase shifts of order of $1$ radian are predicted for a single quantum well using a single sub-picosecond pulse of few $\mu$W power.

The predicted phase nonlinearity can be physically understood as a result of laser-induced dipole effect that dynamically detunes the cavity field with respect to the $1\to2$ intersubband transition, caused by population driven between the first and second excited levels of the anharmonic quantum well spectrum.  
The analytical model is validated numerically using density matrix solutions of the corresponding Lindblad quantum master equation. Notably, the proposed dipole chirping mechanism only occurs for cavity fields that are much shorter lived than the THz dipole resonance (bad cavity limit), as is the case in several nanophotonic setups \cite{Metzger2019,Mann2021}. Moreover, the phase imprinting scheme works best in the weak coupling regime, contrary to conventional photon blockade effects developed for optical cavities QED, which require strong coupling conditions \cite{Birnbaum2005,benea2019electric}. %

Our work demonstrates the feasibility of implementing nonlinear phase operations at THz frequencies using current available nanocavities \cite{Mann2021,Metzger2019,Wilcken2023} and contributes to the development of quantum optics in the high-THz regime \cite{Goulain2023,benea2019electric}, which can enable fundamental studies of cavity quantum electrodynamics \cite{DeLiberato2019,Wang2021}, material and molecular spectroscopy \cite{Kizmann:2022wq,Wilcken2023,Bylinkin:2021}, and controlled chemistry in confined electromagnetic environments \cite{Nagarajan:2021ur,Ahn2023}. Extensions of this work to the analysis of THz and infrared pulses with non-classical field statistics \cite{Waks2004,Zhu2022} could open further possibilities for developing ultrafast quantum information processing at room temperature.

\section{Acknowledgments}

M.A. is funded by ANID - Agencia Nacional de Investigaci\'on y Desarrollo through the Scholarship Programa Doctorado Becas Chile/2018 No. 21181591. 
J.F.T. is supported by ANID - Fondecyt Iniciaci\'on Grant No. 11230679. 
F.H. is funded by ANID - Fondecyt Regular Grant No. 1221420 and the Air Force Office of Scientific Research under award number FA9550-22-1-0245.
A.D. is supported by ANID - Fondecyt Regular 1180558, Fondecyt Grants No. 1231940 and No. 1230586.
All authors also thanks support by ANID - Millennium Science Initiative Program ICN17\_012.

%

%
\bibliographystyle{apsrev4-1}
\bibliography{anhar_phase-v2}

\appendix


\section{MEAN-FIELD APPROACH FOR $N$ IDENTICAL DIPOLES}
\label{app:anharmonic}

The density matrix of the light-matter system $\hat \rho(t)$ evolves according to the quantum master equation in Lindblad form 
\begin{equation}
\frac{\rm d}{\rm dt}\hat \rho = -i[\hat{\mathcal{H}} + \hat H_{\rm d}(t),\hat \rho] + \mathcal{L}_{\kappa}\left[\hat \rho\right] +\mathcal{L}_{\gamma}\left[\hat \rho\right],
\end{equation}
where $\hat{\mathcal{H}}$ is the Hamiltonian of the system in Eq. (\ref{eq:H1}), $\hat{H}_{\mathrm{d}}$ is the time-dependent Hamiltonian of the laser pulse that drives the nanocavity in Eq. (\ref{eq:driving}) and the Lindblad superoperators are given by \cite{Shammah2017,Manzano2020}
\begin{eqnarray}
\mathcal{L}_\kappa[\hat \rho] &=& \frac{\kappa}{2}\left(2\,\hat a\hat \rho\hat a^\dagger - \hat a^\dagger \hat a\, \hat \rho - \hat \rho\,\hat a^\dagger\hat a \right),\label{eq:L kappa}\\
\mathcal{L}_{\gamma}[\hat \rho] &=&\sum_{n=1}^N \frac{\gamma_n}{2}\left(2\,\hat b_n\hat \rho\hat b_n^\dagger - \hat b_n^\dagger \hat b_n\, \hat \rho - \hat \rho\,\hat b_n^\dagger\hat b_n \right).\label{eq:Llocalapp}
\end{eqnarray}
$\kappa$ is the  resonator field decay rate and $\gamma_n$ is the MQW relaxation rate into a local reservoir.
$\hat{a}$ and $\hat{b}_{n}$ are the annihilation operators of the field mode and the $n$-th quantum well, respectively. 

The local operators can be expressed as a linear combination of the collective modes $\hat{B}_{\alpha}$ as
\begin{equation}\label{eq:inverseb}
\hat{b}_n=\frac{1}{\sqrt{N}}\sum_{\alpha=0}^{N-1}\exp\left(-\frac{\ic 2\pi\alpha n}{N}\right)\hat{B}_\alpha.
\end{equation}
Thus, the Lindblad superoperators for the collective modes considering homogeneous MQWs (with $\gamma_n=\gamma$) are given by 
\begin{equation}\label{eq:Llocal2}
\mathcal{L}_{\gamma}[\hat \rho]  =  \frac{\gamma}{2}\sum_{\alpha=0}^{N-1}\left(2\,\hat B_\alpha\hat \rho\hat B_\alpha^\dagger - \{\hat B_\alpha^\dagger \hat B_\alpha\, ,\hat \rho \} \right), 
\end{equation}
where $\{\hat A,\hat B\}$ is the anticommutator between arbitrary operators $\hat A$ and $\hat B$.

%
The exact equations of motion for the field and bright collective matter coherences in the Schr\"odinger picture are given by 
\begin{align}
\label{eq:acubic} \frac{\rm d}{\rm dt}\meanval{\hat{a}}  & = -\left(\frac{\kappa}{2} + i\om{c} \right)\meanval{\hat{a}}  - i \sqrt{N}g  \langle\hat{B}_0\rangle  - i\tilde F_{\rm d}(t),\\
 \frac{\rm d}{{\rm d}t}
\langle{\hat{B}_0}\rangle & = -\left({\gamma}/{2} + i\omega_0\right)\langle{\hat{B}_0}\rangle - i\sqrt{N}g\langle{\hat{a}}\rangle \nonumber\\ \label{eq:B0cubic}
& \quad +i\frac{2U}{\sqrt{N}}\sum_{n=1}^N\langle{ \hat{b}^{\dagger}_n\hat{b}_n\hat{b}_n}\rangle,
\end{align}
with $\tilde{F}_{\rm d}(t) = F_0\varphi(t)e^{-\ic \omega_{\rm d} t}$. We calculate the collective matter coherence via $\langle \hat B_0\rangle =\sum_n{\rm Tr}[\hat b_n\hat \rho(t)]/\sqrt{N}$.
Using Eq. (\ref{eq:inverseb}), the third order term in Eq. (\ref{eq:B0cubic}) is given by an expansion of bright and dark collective operators as
\begin{align}\label{eq:qubic1}
\sum_n \hat{b}_n^\dagger\hat{b}_n\hat{b}_n  & = \frac{1}{N^{\frac{3}{2}}}\sum_{\beta,\eta,\epsilon=0}^{N-1}\left(\sum_n e^{\frac{\ic 2\pi}{N}(\beta-\eta-\epsilon)n}\right)\hat{B}_\beta^\dagger\hat{B}_\eta\hat{B}_\epsilon,\nonumber\\
& = \frac{1}{\sqrt{N}}\sum_{\beta,\eta,\epsilon}(\delta_{\beta-\eta-\epsilon,0} + \delta_{\beta-\eta-\epsilon,-N})\hat{B}_\beta^\dagger\hat{B}_\eta\hat{B}_\epsilon,\nonumber\\
& = \frac{1}{\sqrt{N}}(\hat{B}_0^\dagger\hat{B}_0\hat{B}_0 + \hat{B}_0^\dagger\hat{B}_1\hat{B}_1 +  \ldots)
\end{align} 
Taking the expectation value, the collective third order terms in Eq. (\ref{eq:qubic1}) can be expressed as
\begin{equation}
\langle\hat{B}_\beta^\dagger\hat{B}_\eta\hat{B}_\epsilon\rangle \approx \langle \hat{B}_\beta \rangle^*\langle \hat{B}_\eta \rangle\langle \hat{B}_\epsilon \rangle. 
\end{equation}
The expansion has just one third order term with equal indexes and it is constrained to the bright mode operator ($\beta=\eta=\epsilon=0$), which implies that the remaining terms have at least one dark mode operator, i.e., a nonzero index.
We consider that the laser pulse drives only the nanocavity, which just interacts with the bright mode [see Eq. (\ref{eq:acubic})].  
Hence, considering initial population of the dark modes equal to zero, the dark manifold is completely decoupled from the field $\langle\hat{a}\rangle$ and bright collective mode $\langle\hat{B}_{0}\rangle$. 
The latter is due to the only term in the expansion [Eq. (\ref{eq:qubic1})] that evolves to nonzero values is  $|\langle \hat{B}_0 \rangle|^2\langle \hat{B}_0 \rangle$. 
As a consequence, the evolution of $\langle\hat{B}_{0}(t)\rangle$ reduces to Eq. (\ref{eq:bcoh}).

\section{ADIABATIC ELIMINATION OF THE ANTENNA DYNAMICS }
\label{app:adibaticelimination}
%
%

In the bad cavity limit, we can adiabatically eliminate the dynamics of the single field mode (${\rm d}\meanval{\hat{a}(t)}/{\rm dt}\rightarrow 0$) since $\kappa\gg\gamma$ and $(\kappa-\gamma)/4>\sqrt{N}g$. 
We reduce the equations of motion to a single equation for bright collective matter coherence $\hat{B}_{0}$ which contains the influence of the open cavity mode.
Hence, Eq. (\ref{eq:acubic}) in the rotating frame of the cavity frequency $\omega_{0}$ reduces to 
\begin{equation}
\meanval{\tilde{a}^{\rm ad}(t)} \approx -i\frac{2\sqrt{N}g}{\kappa}\langle\tilde{B}^{\rm ad}_0(t)\rangle - i\frac{2}{\kappa} F_{\rm d}(t), 
\end{equation}
and solving for $\langle\hat{B}_{0}\rangle$ 
\begin{align}
\frac{\rm d}{\rm dt}\langle\tilde{B}^{\rm ad}_0(t)\rangle & = -\frac{\tilde{\gamma}}{2}\langle\tilde{B}^{\rm ad}_0(t)\rangle - \frac{2\sqrt{N}g}{\kappa}  F_{\rm d}(t) \nonumber\\ \label{eq:bss}
& \quad + i\frac{2U}{N}|\langle\tilde{B}^{\rm ad}_0(t)\rangle|^2\langle\tilde{B}^{\rm ad}_0(t)\rangle, 
\end{align}
with the renormalized decay rate of the dipole coherence $\tilde{\gamma} =\gamma(1+4Ng^2/\kappa\gamma)$, which is commonly known as the Purcell factor \cite{herrera2022}.

Equation  (\ref{eq:bss}) with $F_{\mathrm{d}}(t)=0$ is known in non-linear hydrodynamics as the Stuart-Landau oscillator equation \cite{Stuart1958,Panteley2015}. 
The laser pulse at a given time $t=t_{\mathrm{off}}$ turns off and Eq. (\ref{eq:bss}) can be solved analytically by a slow variation of the dipole coherence in polar form as $\langle\tilde{B}^{\rm ad}_0(t)\rangle = |\langle\tilde{B}^{\rm ad}_0(t)\rangle|e^{i\phi(t)}$. Thus, the equations of motion for the amplitude and phase can be written as
\begin{align}
\frac{\rm d}{\rm dt}|\langle\tilde{B}^{\rm ad}_0(t)\rangle|& =- \frac{\tilde{\gamma}}{2}|\langle\tilde{B}^{\rm ad}_0(t)\rangle| \\
\frac{\rm d}{\rm dt}\phi(t) & = 2\frac{U}{N}|\langle\tilde{B}^{\rm ad}_0(t)\rangle|^2. 
\end{align}
where their corresponding solutions are given by
\begin{align}\label{eq:ampb0_fd0}
|\langle \tilde{B}_0^{\rm ad}(t)\rangle| & = B_{\rm off} e^{-\frac{\tilde{\gamma}}{2}(t-t_{\rm off})}\\ \label{eq:ampphi_fd0}
\phi(t) & = \phi_{\rm off} + \frac{2U B_{\rm off}^2}{N\tilde{\gamma}}\left\{1-e^{-\tilde{\gamma}(t-t_{\rm off})}\right\}.
\end{align}
with $B_{\rm off}=|\langle \tilde{B}_0^{\rm ad} (t_{\rm off})\rangle|$ and $\phi_{\rm off}=\phi(t_{\rm off})$.
The dipole coherence in the rotating frame of the laser evolves as
\begin{align}
\label{eq: ad nonlinearB0}
\langle\hat{B}_{0}^{\rm ad}(t)\rangle & = \langle\hat{B}_{0}^{\rm ad}(t_{\rm off})\rangle e^{-
\frac{\tilde{\gamma}}{2}
(t-t_{\rm off})}e^{-i
\omega_0 (t-t_{\rm off})}\nonumber\\
& \quad \times 
\exp[i\Delta\phi_{\rm ss}(1-\exp[-\tilde{\gamma}(t-t_{\rm off})])], 
\end{align}
where $\Delta\phi_{\rm ss} = \phi_{\rm ss} - \phi_{\rm off}$ and $\langle\hat{B}_{0}^{\rm ad}(t_{\rm off})\rangle = B_{\rm off}e^{i\phi_{\rm off}}$.
The exponential that depends on the relative phase $\Delta\phi$ in Eq. (\ref{eq: ad nonlinearB0}) evidences the nonlinear contributions, instead of the solution with harmonic MQWs or in the weak driving regime.
To clarify, the analogous solution of the dipole coherence with $U=0$ [$\hat{B}_{0,{\rm L}}^{\rm ad}(t)$] for $t\geq t_{\rm off}$ is given by 
\begin{equation}
\label{eq: ad linearB0}
\langle\hat{B}_{0,{\rm L}}^{\rm ad}(t)\rangle  = - \beta(T)\frac{2\sqrt{N}gF_0}{\kappa}e^{-(\tilde{\gamma}/2 + i\omega_0)t} 
\end{equation}
where the factor $\beta(T)$ depends on the envelope functional shape, and the stationary phase $\Delta\phi_{\rm ss}=0$ due to the system evolves with a constant phase $\phi_{\rm ss}=\phi_{\rm off}$.

In the case of the phase, Eq. (\ref{eq:ampphi_fd0}) describes a stationary phase $\phi(t)=\phi_{\rm ss}$ in the long time regime ($t\gg t_{\mathrm{off}}$), which is given by 
\begin{equation}\label{eq:phss1}
\phi_{\rm ss}= \phi_{\rm off} + \frac{2U B_{\rm off}^2}{N\tilde{\gamma}}.
\end{equation}
Note that the expression is quadratic respect to amplitude $B_{\mathrm{off}}$ and constant for harmonic MQWs ($U=0$).

\subsection{Relation between the dipole and cavity phase shifts}

We measure the free-induction decay signal in the laboratory, which is related with the field mode coherence $\langle\hat{a}\rangle$. Here, we connect the phase shift that can be obtained from experiments with the phase from the dipole coherence. 
We define the relative nonlinear phase shift in frequency domain at $\omega_0$ in terms of the Fourier transform of the cavity coherence, $\langle \hat{a}(\omega)\rangle = \mathcal{F}[\langle \hat{a}(t)\rangle](\omega)$, as
\begin{equation}\label{eq:phaseshift2}
\Delta\Phi(\omega_0)  = 
 \Phi(\omega_0)-\Phi_{\rm harm}(\omega_0),
\end{equation}
where,
\begin{equation}
\Phi(\omega) = \arctan\left(
\frac{{\rm Im}[\langle \hat{a}(\omega)\rangle]}{{\rm Re}[\langle \hat{a}(\omega)\rangle]}
\right)
\end{equation}
and $\Phi_{\rm harm} = \lim_{F_0/\kappa\ll 1}\Phi(\omega_0)$.
The latter is valid since the response of the anharmonic dipole oscillator under weak driving conditions $F_0/\kappa\ll 1$ or in the limit of negligible anharmonicity $U/\gamma\to 0$ is equivalent to the linear response, as it is shown in Eq. (\ref{eq:phss1}). 
Assuming the same $t_{\mathrm{off}}$ for the cavity and dipole coherences, the equations of motion for the field coherence and phase in frequency domain are given by
\begin{align}
\langle \hat{a}(\omega) \rangle & = -ig_N\langle \hat{B}_0(\omega) \rangle \frac{1}{\kappa/2 - i(\omega - \omega_{\rm c})}, \\
\label{eq: a phase}
\Phi(\omega_0) & = \Phi^{(0)}(\omega_0) + \arctan\left(
\frac{\omega_0 - \omega_{\rm c}}{\kappa/2}
\right) - \frac{\pi}{2}, 
\end{align}
with
\begin{equation}
\nonumber\Phi^{(0)}(\omega_0) = \arctan\left(
\frac{{\rm Im}[\langle \hat{B}_0(\omega_0) \rangle]}{{\rm Re}[\langle \hat{B}_0(\omega_0) \rangle]}
\right).
\end{equation}
The second and third term in Eq. (\ref{eq: a phase}) are independent on anharmonicity parameter $U/\kappa$ and driving strength $F_0/\kappa$. 
Thus, in analogy with Eq. (\ref{eq:phaseshift2}), i.e., considering that the relative phase shift is given in terms of the linear response and nonlinear contributions, we can write $\Delta\Phi(\omega_{0})$ as a function of the dipole coherence instead of the field mode response as 
\begin{equation}\label{eq:phaseshift3}
\Delta\Phi(\omega_0)  =\Delta\Phi^{(0)}(\omega_0) 
 =  \Phi^{(0)}(\omega_0)-
 \Phi_{\rm harm}^{(0)}(\omega_0),
\end{equation}
where $ \Phi_{\rm harm}^{(0)}(\omega_0) = \lim_{F_0/\kappa\ll 1}\Phi^{(0)}(\omega_0)$.

\subsection{Nonlinear phase shift ansatz for an arbitrary driving pulse}

We introduce an ansatz for the relative phase since the amplitude $|B_{\rm off}|$ cannot be defined for general driving pulses. 
We define the nonlinear phase shift $\Delta\Phi$ at frequency $\omega_{0}$ as 
\begin{equation}
\label{eq:phenomelogicalphase}\Delta\Phi(\omega_0)  =\alpha
\frac{2U}{N\tilde{\gamma}}\left(
\frac{F_0}{\kappa}
\right)^2,
\end{equation}
where $\alpha$ is a phenomenological parameter to be explored.
The definition in Eq. (\ref{eq:phenomelogicalphase}) is possible considering that the squared amplitude of the dipole coherence [Eq. (\ref{eq:ampb0_fd0})] and the stationary phase [Eq. (\ref{eq:phss1})] grow proportional to the square of the driving strength for the ratio $F_0/\kappa < 1$.
Further, numerical results in Fig. \ref{fig:freqdomain} suggest the quadratic dependence.

\section{MEAN-FIELD CHIRPING MODEL FOR TWO ASYMMETRIC QUANTUM WELLS}
\label{app:asymmetric}

From the local quantum master equation, the mean-field equations of motion for the coherences of the inhomogeneous quantum wells ($\hat{b}_{1}$ and $\hat{b}_{2}$) and field mode $\hat{a}$ are given by 
\begin{align}
\frac{\rm d}{\rm dt}\langle \hat{a} \rangle & = -\left(\frac{\kappa}{2}+i\omega_{\rm c}\right)\langle \hat{a} \rangle - i g(\langle \hat{b}_1\rangle + \langle \hat{b}_2\rangle) - i\tilde{F}_{\rm d}(t),\\
\frac{\rm d}{\rm dt}\langle \hat{b}_1 \rangle & =  -\left(\frac{\gamma_1}{2}+i\omega_{\rm 1}\right)\langle \hat{b}_1 \rangle - i g\langle \hat{a}\rangle +i2U|\langle \hat{b}_1 \rangle|^2\langle \hat{b}_1 \rangle,  \\
\frac{\rm d}{\rm dt}\langle \hat{b}_2 \rangle & =  -\left(\frac{\gamma_2}{2}+i\omega_{\rm 2}\right)\langle \hat{b}_2 \rangle - i g\langle \hat{a}\rangle  +i2U|\langle \hat{b}_2 \rangle|^2\langle \hat{b}_2 \rangle. 
\end{align}
We set equal light-matter coupling strengths $g=g_{1}=g_{2}$ and anharmonicity parameters $U=U_{1}=U_{2}$. 
By replacing the bright $\hat{B}_0=(\hat{b}_1 + \hat{b}_2)/\sqrt{2}$ and dark $\hat{B}_1=(-\hat{b}_1 + \hat{b}_2)/\sqrt{2}$ modes, we obtain
\begin{align}
\frac{\rm d}{\rm dt}\langle \hat{a} \rangle & = -\left(\frac{\kappa}{2}+i\omega_{\rm c}\right)\langle \hat{a} \rangle - i g\langle \hat{B}_0\rangle  - i\tilde{F}_{\rm d}(t),\\
\begin{split}
\frac{\rm d}{\rm dt}\langle \hat{B}_0 \rangle & =  -\left(\frac{\bar{\gamma}}{2}+i\bar{\omega}(t)\right)\langle \hat{B}_0 \rangle - \left(\frac{\Delta\gamma}{2} + i\Delta\omega(t)\right)\langle \hat{B}_1 \rangle\\
& \quad - i g\langle \hat{a}\rangle, 
\end{split} \\
\begin{split}
\frac{\rm d}{\rm dt}\langle \hat{B}_1 \rangle & =  -\left(\frac{\bar{\gamma}}{2}+i\bar{\omega}(t)\right)\langle \hat{B}_1 \rangle - \left(\frac{\Delta\gamma}{2} + i\Delta\omega(t) \right)\langle \hat{B}_0 \rangle,
\end{split}
\end{align}
with 
\begin{align*} 
\bar{\omega}(t) & = \bar{\omega} - U\left(|\langle \hat{B}_0(t) \rangle|^2+|\langle \hat{B}_1(t) \rangle|^2\right),
\end{align*}
and 
\begin{align*} 
 \Delta\omega(t) & = \Delta\omega - 2U{\rm Re}[\langle \hat{B}_0(t) \rangle^*\langle \hat{B}_1 (t)\rangle].
\end{align*}
where $\bar{\zeta}=(\zeta_1+\zeta_2)/2$ and $\Delta\zeta = (\zeta_2 - \zeta_1)/2$, with $\zeta=\{\gamma,\omega\}$, are the average and mismatch values.

\end{document}